\documentclass[rnote,letterpaper]{aa}
\usepackage{graphicx}
\usepackage{txfonts}
\begin{document}
\title{The Transition between Star Clusters and Dwarf Galaxies}

\subtitle{On the existence of a mass--radius relation for star clusters
of masses $>10^7 M_\odot$: are these objects formed in
mergers of stellar systems? }

\author{M. Kissler-Patig\inst{1}
   \and
        A. Jord\'an\inst{1,2}
   \and
        N. Bastian\inst{3}
       }

   \offprints{M. Kissler-Patig}

   \institute{European Southern Observatory, Karl-Schwarzschild-Str.2,
     85748 Garching, Germany\\
     \email{mkissler@eso.org, ajordan@eso.org}     
     \and
     Astrophysics, Denys Wilkinson Building, University of Oxford,
     1 Keble Road, Oxford, OX1 3RH, UK
     \and
     Department of Physics and Astronomy, University College
     London, Gower Street, London, WC1E 6BT, UK\\
     \email{bastian@star.ucl.ac.uk}
   }

   \date{Received ... ; accepted ... }

   \abstract
{At which masses does the regime of globular clusters end and the one of
dwarf galaxies begin? And what separates these two classes of hot
stellar systems?}
{We examine to what extend very massive ($>10^7 M_\odot$) young
star clusters are similar to their lower mass counter parts and to which
degree they resemble other objects in their mass regime
(dwarf--globular transition objects (DGTOs), ultra compact dwarf
galaxies (UCDs), galaxy nuclei)}
{The comparison is performed by placing the recently observed very massive 
young clusters onto known scaling relation defined by globular
clusters (with typical masses $\la10^6 M_\odot$) and/or by hot stellar
systems with sizes up to those of giant galaxies.}
{The very massive ($\ga10^{6.5-7} M_\odot$) young clusters seem to show a  
mass--radius relation compatible with the one defined by hot stellar systems 
of galaxy mass. This, in turn, can explain their location on the other scaling 
relations investigated. It contrasts with the behaviour of the less massive young 
clusters and of globular clusters, which do not exhibit any mass-radius
relation.  However,  the behaviour of the most massive clusters is similar to 
that of  most other objects in that mass regime ($10^6-10^8 M_\odot$).}
{We show that the properties of young massive clusters are compatible with
other objects in the same mass regime such as DGTOs/UCDs. They present a 
possible direct avenue of formation for those objects, which does not require 
the transformation of a previously existing stellar system.
Simulations and observations support the possibility of the  
formation of such very massive young clusters by early mergers of 
lower mass stellar clusters, which could explain the emergence of
a mass--radius relation.}
   \keywords{star clusters -- dwarf galaxies}

   \maketitle
%

\section{The transition region between massive star clusters
and low-mass galaxies}

In the past decade, observations have slowly filled the previous gap in the
mass distribution of compact objects between globular clusters in the Milky 
Way ($\la 10^6 M_{\odot}$) and compact low-mass galaxies such as M32 (several 
$10^8 M_{\odot}$). This has triggered many questions about the nature of such
objects, which lie in the mass range $10^6$ to $10^8 M_\odot$.

Attention was first drawn to these objects by the discovery of very massive 
($\ga10^6 M_{\odot}$), compact stellar systems in Fornax (Minniti et
al.~1998, Hilker et al.~1999a,b, Drinkwater et al.~2000), later baptised 
`Ultra Compact Dwarf' (UCD)
galaxies. The origin of these objects is still hotly debated in the
literature: they have been proposed to be the high mass end of the
globular cluster population, the nuclei of stripped galaxies,
merged star clusters (`stellar superclusters') or a new class of compact dwarf 
galaxies (Hilker et al.~1999a,b, 
Drinkwater et al.~2000, Phillipps et al.~2001, Fellhauer \& Kroupa 2002,
Bekki et al.~2003, 2004, Drinkwater et al.~2003, Maraston et al.~2004, 
Mieske et al.~2004, Ha\c segan et al.~2005). 
Ha\c segan et al.~(2005) introduced the term ``Dwarf-Globular Transition 
Objects'' (DGTOs) in order to emphasize the ambiguity of their
classification.

The fundamental open question is the formation process of these
objects. To answer this question, their location on various scaling
relations has been investigated and compared to the loci of other
known objects in that mass regime (such as massive globular clusters, 
nuclei of dwarf galaxies, nuclear clusters in bulge-less spirals, 
simulated merged clusters; see above references and Walcher et al.~2005).

Young massive clusters were first placed on $\sigma$ against $M_V$
scaling relation in Kissler-Patig (2004) and in the $\kappa$--plane in
Maraston et al.~(2004), and it was noticed that the
the most massive cluster departed from the globular cluster relation in the 
direction of UCDs (see also de Grijs et al.~2005 for an extensive
discussion). However, as discussed also in Bastian et al.~(2005a),
such relations are problematic as they require a
correction of the absolute magnitude $M_V$ for evolution (prone to errors in 
the distance, the age and the evolutionary model) which complicates the 
analysis.

Ha\c segan et al.~(2005) suspect a break in the scaling relations of
star clusters and galaxies around $10^6 M_{\odot}$.  The authors suggest 
that above that mass, DGTOs/UCDs appear to split in two 
groups. Some DGTOs/UCDs follow an extrapolation of the globular
cluster scaling relations to high masses: they are considered to be globular 
clusters with unusually high mass and luminosity. 
Others fall along the galaxy scaling relations: they are viewed as prime
UCD candidates.

Geha et al.~(2002) put dE,N nuclei on scaling relations and find them to
fall in the range spanned by globular clusters, although slightly offset in
mass. No direct comparison with DGTOs/UCDs was made. 

Walcher et al.~(2005) notice that nuclear clusters (thought to grow
by continuous mass accretion), when put on scaling relations, are in general 
more compact then UCDs.

Fellhauer \& Kroupa (2002) and Bekki et al.~(2004) argue that the
properties of the simulated product of multiple merging of star clusters
reproduces well the properties of UCDs/DGTOs. 

In this contribution, we add a piece to the puzzle by placing recently
observed {\it young massive star clusters} with masses $\ga10^7 M_{\odot}$ on the 
various scaling relations. This allows us to shed new light on the 
origin of the objects in this regime. It also shows that objects with the
properties of UCDs/DGTOs can be formed directly as a consequence of
star cluster formation. Thus, their formation does not seem 
to necessarily require the transformation of a parent stellar system
such as a dwarf galaxy.


\begin{figure}
\centering
\includegraphics[width=9cm]{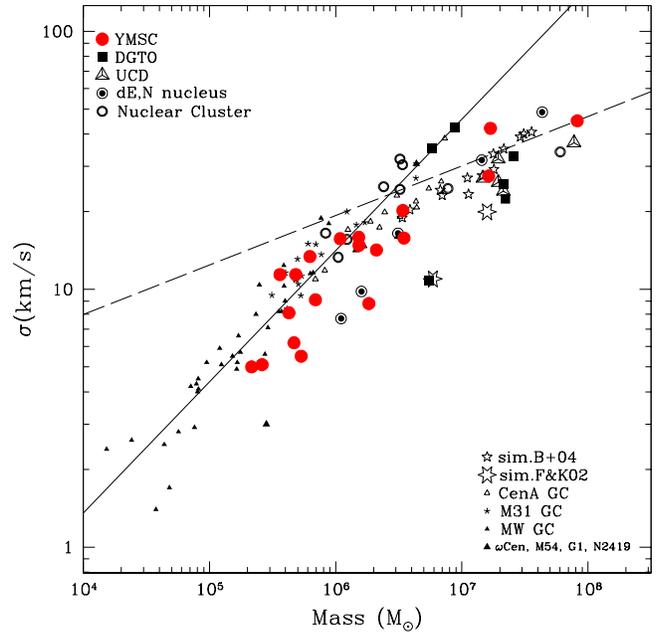}
\caption{Scaling relations for low-mass, hot stellar systems:
line-of-sight velocity dispersion plotted against total mass. For
details on the objects plotted, see text. The solid and dashed lines show the
fitted relations for globular clusters and elliptical galaxies,
respectively.}
\label{sigmass}
\end{figure}
\begin{figure}
\centering
\includegraphics[width=9cm]{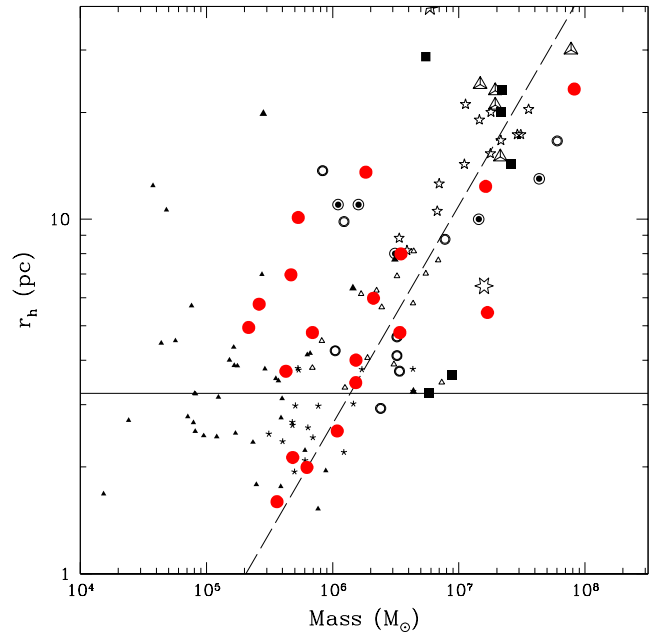}
\caption{Scaling relations for low-mass, hot stellar systems:
half-light radius plotted against total mass. For
details on the objects plotted, see text (symbols as in
Fig.~\ref{sigmass}). The dashed line shows the
fitted relation for elliptical galaxies, while the solid lines indicates
the median for Galactic globular clusters ($r_{h}$=3.2 pc) that do not
follow a mass-radius relation.}
\label{radmass}
\end{figure}

\begin{figure}
\centering
\includegraphics[width=9cm]{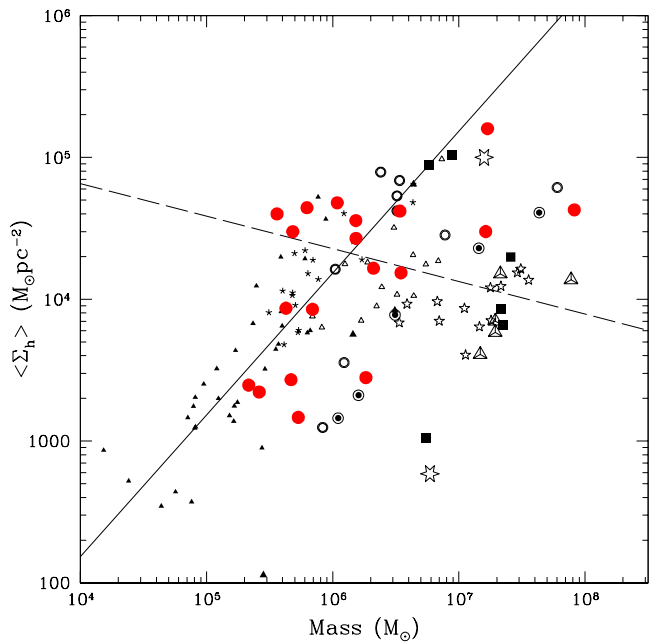}
\caption{Scaling relations for low-mass, hot stellar systems:
mean mass density within the half-light radius plotted against total mass. For
details on the objects plotted, see text (symbols as in
Fig.~\ref{sigmass}). The solid and dashed lines show the
fitted relations for globular clusters and elliptical galaxies,
respectively.}
\label{mumass}
\end{figure}
%

\section{Young star clusters with mass of $10^6 M_\odot$ and above}

We are searching for counter-parts of UCDs/DGTOs, i.e.~compact objects with
masses $>10^6 M_{\odot}$ (ideally $>10^7 M_{\odot}$) {\it of known origin}. 
Massive star/globular clusters are the first objects that come to mind, but 
star clusters with masses $>10^7 M_{\odot}$ are not known in the Local Group. 
Moreover, old objects (e.g.~as $\omega$ Centauri) have a nature/origin that is 
still under debate (e.g.~Hilker et al.~2004 and reference therein). 

However, some {\it young} massive star clusters found in major star forming 
regions (typically galaxy collisions) are known to have (photometrically 
determined) masses in the regime of interest. The advantage of young
objects is that they do not have much of a past. Their properties reflect
their short period of formation. Their age is known, so that even if
they are made up by merging fragments or clusters, all of them formed in
the same starburst. Furthermore, young massive clusters (as opposed to
e.g.~a nucleus of a dwarf galaxy stripped of its envelope) are not
expected to have any significant dark matter.

Recently, a handful of such system have had their masses determined
{\it dynamically} (Maraston et al.~2004, Bastian et al.~2005a). They are prime
candidates to be compared with UCDs. Maraston et al.~(2004) did so for
the most massive of them (NGC7252:W3), and came to the conclusion that it has
remarkably similar properties to the mysterious UCDs. The case which was made 
for one object may have failed to make a general point, but is reinforced
below with the addition of new objects from Bastian et al.~(2005a).


\section{Young massive star clusters on the scaling relations of low-mass, 
hot stellar systems}

\subsection{Placing the young massive star clusters on the scaling
relations}

Scaling relations are recognized as prime tools to understand the
evolution of hot stellar systems 
(e.g., Djorgovski \& Davis 1987; McLaughlin 2000). 
Given that we aim to compare young systems
with old ones, we chose to avoid any relation that involves the Mass-to-Light
ratio. Although we think that the `aging' of the young star
clusters is well understood, we prefer not to add this uncertainty to our analysis.
We note that Maraston et al.~(2004) and 
Bastian et al.~(2005a) show convincingly that
the systems will evolve into stellar populations with `normal' M/L ratios 
similar to DGTOs/UCDs, i.e.~between 2 and 6, as expected for old metal-rich
system not significantly dominated by dark matter.

Thus, we exclude the Faber-Jackson relation (Faber \& Jackson 1976) as
well as the $\kappa$--space (Bender et al.~1992, Burstein et al.~1997)
to focus on relations between mass and, in turn, velocity dispersion,
half-light radius, and mean mass density within the half-light radius 
(see Ha\c segan et al.~2005 for a detailed discussion of these relations).

In order to compare the young clusters with the old objects on
these relations, we need to estimate their evolution in the above
quantities. Literature on the dynamics of star clusters and their
evolution is extensive and spans the last century. We mention here only
some of the most recent N-body simulations that focus on star clusters with
high masses (Fellhauer \& Kroupa 2005, Baumgardt et al.~2004).
The simulations agree on the fact that after the first few $10^6$ years, massive
clusters evolve only slowly in mass, velocity dispersion and radius. For
example, Fellhauer \& Kroupa (2005, see also Fellhauer \& Kroupa 2002) 
specifically tried to reproduce the cluster NGC7252:W3. After $\sim 300$ Myr
(a lower limit for the cluster age), the velocity dispersion decreases by
10-20\% over the next several Gyr, while the mass drops by $\sim$30\%, and the
effective radius reacts by rising to $\sim$ 10 pc. Note that the
relatively small simulated effective radius is at odds with the observations. 
Fellhauer \& Kroupa argue the observed larger radius
could be explained if NGC7252:W3 was seen at a the time of a late merging event.
The clusters NGC7253:W30 ($R_{\rm eff}\sim 9$ pc) and NGC1316:G114 ($R_{\rm
eff}\sim 4$ pc) are better matched by the simulations.

The cluster of most interest in our context are the most massive known 
(NGC7252:W3, NGC7252:W30, NGC1316:G114). They have ages between 500 Myr and 3 
Gyr and their mass, radius and velocity dispersion are thus likely to evolve by
modest factors as discussed above. We do not attempt to
correct for evolution in our comparison but keep this uncertainty in
mind when reaching conclusions.

We used the young star-cluster data as tabulated in Maraston et 
al.~(2004) and Bastian et al.~(2005a, including the literature collection of 
their table 5 for less massive young star clusters). The projected
effective radius ($R_{\rm eff}$) was transformed into an unprojected 
half-light radius ($r_{h}$) according to $r_{h}=1.3 R_{\rm eff}$ (Spitzer 1987). 
The mass ($M$) was computed according to $M=\eta \sigma^2 {r}_{h} / G$, 
where $\sigma$ is the
line-of-sight velocity dispersion, $G$ the gravitational constant, and
$\eta=7.5$ (see Boily et al.~2005 and references therein for discussions
on $\eta$) \footnote{with $\sigma$ in km s$^{-1}$ and $r_{h}$ in pc,
this results in $M/M_\odot=1744\cdot \sigma^2 {r}_{h}$}. The mean
mass density within the half-light radius, $\langle \Sigma_{h} \rangle$, is
derived from the two above quantities 
(as $\langle \Sigma_{h} \rangle= 0.5 M/(\pi {R}_{\rm eff}^2)$).

The data for young massive star clusters ({\it solid circles}) are plotted on 
three scaling relations in Figs.~\ref{sigmass}, \ref{radmass} and
\ref{mumass}, together with the data for other low-mass, hot stellar systems.
Specifically, we show globular clusters belonging to the Milky Way, 
M31 (McLaughlin \& van der Marel 2005), and NGC 5128 (Martini \& Ho
2004, Harris et al.~2002), nuclei of dE,N (Geha et al.~2002), UCDs
(Drinkwater et al.~2003), DGTOs (Ha\c segan et al.~2005), nuclear
clusters (Walcher et al.~2005), as well as 
simulation of multiple merged star clusters (Fellhauer \& Kroupa 2002,
2005, Bekki et al.~2004, for the latter only selected data points that cover 
the range of properties span by the simulations). We also plot the fitted 
scaling relations for globular clusters ({\it dashed line}) and elliptical
galaxies ({\it solid line}) as derived in Ha\c segan et al.~(2005).

Note that the three scaling relations are not independent as they
plot, in the above order, $\sigma^2$ against $\sigma^2 r_h$,
$r_h$ against $\sigma^2 r_h$ and $\sigma^2 / r_h$ against $\sigma^2 r_h$.
Yet, they exhibit different enough views on these quantities that they
are worth being considered individually.

\subsection{Young massive star clusters: where do they fit?}

We review below where the young massive star clusters fall on the 
three scaling relations.

In Fig.~\ref{sigmass}, the young massive star clusters follow nicely the
relation spanned by the globular cluster for masses $<10^6 M_\odot$ to
then continue at higher masses along the sequence defined by elliptical 
galaxies. Thus, while the lower-mass star clusters are indistinguishable 
from globular clusters, the higher-mass ones consolidate a sequence on which 
also the most massive DGTOs, nuclear clusters, UCDs, dE,N nuclei fall. The
$\sigma$--mass relation `bends over' for masses greater than $\sim
3\times 10^6 M_\odot$. This can be understood as a consequence of the
mass--radius relation (see below) that appears to hold for objects above
this mass: more massive clusters will be less dense, which in turn will
lead to a lower velocity dispersion. The most massive young star clusters are
indistinguishable in this relation from UCDs/DGTOs, dE,N nuclei and
nuclear clusters.

Fig.~\ref{radmass} shows a relation between radius and total mass
for the young massive star clusters. They fall on the relation defined
by elliptical galaxies, on which also DGTOs/UCDs, dE,N nuclei and
nuclear clusters lie. Such a relation is not observed for globular clusters 
(McLaughlin 2000, Jord\'an et~al. 2005), nor for young massive 
clusters with masses below $10^5 M_\odot$
(Zepf et al.~1999, Larsen 2004, Bastian et al.~2005b). 
Other trends are apparent, e.g.~the young
massive star clusters with $<10^6 M_\odot$ might be systematically
slightly larger at a given mass then defined by the overall relation.
Uncertainties in the mass and radius determination make any firm statements
difficult, though.

The third scaling relation (Fig.~\ref{mumass}) 
exhibits a similar results as Fig.~\ref{sigmass} (as it essentially shows
$\sigma^2 / r_h$ against $\sigma^2 r_h$ instead of $\sigma^2$
against $\sigma^2 r_h$). Interestingly, the most massive, young star clusters
range among the objects with the highest mass densities at a given mass. 
NGC1316:G114 even exhibits the highest mass density inside a half-light
radius ever observed among low-mass hot stellar systems ($\langle
\Sigma_{h} \rangle > 10^5 M_\odot pc^{-2}$). The large
scatter prevents from defining a clear trend, but again the young
massive star cluster appear to `bend over' and to follow a shallower
relation than the one defined by globular clusters.

In all three relations, the simulated products of multiple merging of
star clusters (Bekki et al.~2004, Fellhauer \& Kroupa 2005) reproduce very 
well the most massive, young star clusters. The Fellhauer \& Kroupa
(2005) simulation produce quite compact star clusters, and thus are unable to 
fit NGC7252:W3 and NGC7252:W30, but match well the observed properties of
NGC1316:G114, which the simulations of Bekki et al.~(2004) fail to reproduce 
well.

We note, finally, that the (seven) young massive clusters lying well
{\it below} the mass--surface density relation of {\it galaxies} 
(see Fig.~\ref{mumass}, $\langle \Sigma_{h} \rangle < 10^4 M_\odot
pc^{-2}$), are also the ones that lie the furthest below the
mass--$\sigma$ relation of galaxies, and the ones to the far left of the
mass-radius relation.


\section{Discussion and Conclusions}

The newest mass measurements of young clusters with masses greater than 
$10^7 M_\odot$ show that these objects 
overlap in the scaling relations with DGTOs/UCDs and other objects in
that mass regime. In particular, the most massive young clusters 
seem to follow the same mass--radius relation as DGTOs/UCDs and elliptical 
galaxies.
{\it This suggests that DGTOs/UCDs are compatible with having the same
nature/origin as the most massive young clusters.} An open question is the 
ability for the evolved products of these massive young clusters to reproduce
the high mass-to-light ratios observed for DGTOs (Ha\c segan et al.~2005).

We can then ask: what is the formation process for these most massive
young clusters? As mentioned already above, independent simulations (Bekki 
et al.~2004, Fellhauer \& Kroupa 2005), as well as recent observations
(Larsen et al.~2002, Minniti et al.~2004, Bastian et al.~2005c) point towards 
the idea that these objects could be products of early star cluster mergers, 
occurring in the first tens to hundred Myr. Our results are consistent with 
this hypothesis.

Note that {\it late} mergers of star/globular clusters are not excluded as
an alternative formation process for very massive star clusters.  Oh \& Lin's 
(2000) simulations show that old star clusters can, through orbital decay, 
sink into the center of a dwarf galaxy and assemble to form a very 
massive star cluster. While it could potentially apply to (some?) 
DGTOs/UCDs, it cannot explain {\it young} massive clusters such as
discussed here. 

Are the most massive young clusters a distinct class of objects with
respect to globular or lower mass young clusters? At face value, the scaling
relations appear to differ. In particular, globular clusters and young
massive clusters with masses of less than $10^5 M_\odot$ do not appear to 
follow a mass--radius relation (which in turn, when combined with the virial
theorem, would explain why the $\sigma$--mass and $\langle \Sigma_{\rm
h}\rangle$--mass 
relations `bend over'). 
The study of star cluster complexes and their associated giant molecular
clouds (Bastian et al.~2005c) showed that the emergence of a mass--radius 
relation seems to occur at scales between individual star clusters and cluster
complexes. The complexes show a similar relation as their parent giant
molecular clouds. This would strengthen the assumption that the most massive 
star clusters, and potentially some DGTOs/UCDs, are associated with star 
cluster complexes, i.e.~star cluster merger events. 

As an alternative to the above scenario, one could speculate that all star 
cluster form with a primordial mass--radius relation, but only the most 
massive star clusters are able to retain it against processes that would erase 
it (cf.~Ashman \& Zepf 2001, Bastian et al.~2005c). But given the lack of
theoretical and observational support for this scenario, 
we currently favor the first hypothesis: objects in the mass range
$10^6\leq M \leq 10^8 M_\odot$ are likely to be
the lowest mass structures resulting from merger of stellar systems, with
the merging process being at the origin of a mass-radius relation.  
If the most massive star clusters indeed form by mergers and
the typical globular clusters not, the two populations must overlap in
some mass regime ($10^5 - 10^6 M_\odot$?). 
The caveat is that the scenario cannot answer the question why globular 
clusters do not show a mass--radius relation, in contrast to the molecular
clouds in which they are thought to be formed.
But this scenario, in which young massive star clusters are the product
of star cluster merger events,
would explain their presence at the lowest mass end of the galaxy scaling
relations extending from the dwarf galaxy regime to giant ellipticals.


\begin{acknowledgements}
      
We thank Monica Ha\c segan and Carl Jakob Walcher for providing
us with electronic compilations of their data. We enjoyed 
useful discussions with Mark Gieles. The referee, S.Phillipps, is
thanked for useful comments.

\end{acknowledgements}

\end{document}